\documentclass[twoside,12pt]{article} 
\usepackage{epsfig} 
 
\def\Journal#1#2#3#4{{#1} {#2} (#4) #3 } 
\def\AA{{\em Astron. \& Astrophys.}}       \def\APJ{{\em Astrophys. J.}}    
    
 \def\PLB{{\em Phys. Lett.} B}
   \def\PRL{\em
  Phys. Rev. Lett.}    \def\PREP{\em Phys.
  Rep.}   \def\PRD{{\em Phys. Rev.} D}
\def\PRC{{\em Phys. Rev.} C} 
  \def\SC{{\em Science}}
        
\def\arx{{\em arXiv:}}   
  
\newcommand{\be}{\begin{equation}} \newcommand{\ee}{\end{equation}}
\newcommand{\bea}{\begin{eqnarray}} \newcommand{\eea}{\end{eqnarray}}
\begin{document} 
 
\title{ \vspace{1cm} 
Modern compact star observations and the quark matter EoS}

\author{T. Kl\"ahn$^{1,2}$
\\ 
$^1$Institute of Physics, Rostock University, D-18051 Rostock, Germany\\ 
$^2$Gesellschaft f\"ur Schwerionenforschung mbH, D-64291 Darmstadt, Germany} 
\maketitle 
\begin{abstract} 
A hybrid equation of state (EoS) for dense
matter is presented that satisfies phenomenological constraints
from modern compact star (CS) observations which indicate
high maximum masses ($M\sim 2 M_\odot$) and large radii
($R> 12$ km).
The corresponding isospin symmetric EoS
is consistent with flow data analyses of heavy-ion collisions.
The transition from nuclear to two-flavor color superconducting
quark matter at $n_{\rm crit}\sim 0.55$ fm$^{-3}$ 
is almost a crossover.
\end{abstract}
Constraints on the high density EoS emerge
from analyses of the elliptic flow in heavy ion collisions (HICs)
and astrophysical data regarding the mass and
mass-radius relations of neutron stars.
Altogether they form a valuable benchmark
for the reliability of a given model EoS as exemplified
for a set of modern nuclear EsoS in \cite{Klahn:2006ir} 
with the result that every  EoS fulfills some of these constraints 
but none could satisfy all of them.
In the following a subset of these constraints
is applied in order to investigate
the compatibility of the presence of a quark matter core
in the neutron stars interior with 
present phenomenological findings.

The stiffness of the symmetric EoS is
limited by analyses of the elliptic flow
in HICs \cite{Danielewicz:2002pu} as
indicated by the hatched area in the right panel of Fig.~1.
The left panel of Fig.~1 illustrates the astrophysical constraints
applied in this paper.
The object PSR J0751+1807 gives a lower limit on the maximum NS mass
of $\approx 1.9 M_\odot$ \cite{NiSp05},
whereas the thermal emission of RX J1856-3754 provides a lower limit
in the mass-radius plane, which suggests minimal radii of $R > 12 $km
for expected NS masses \cite{Trumper:2003we}.
These latter three constraints 
form a {\it minimum requirement} on a reliable hybrid EoS.

It has been questioned whether they could be compatible with a phase transition
to quark matter, which is expected to sufficiently soften the EoS and thus lowering the
maximum NS mass. 
In \cite{Alford:2006vz} a set of counter examples has been provided.
Here we discuss a hybrid EoS where the quark matter phase is described by 
a color superconducting 3-flavor NJL model \cite{Blaschke:2005uj}
augmented by a selfconsistent vector meanfield responsible for stiffening
the EoS \cite{Klahn:2006iw}.
The nuclear matter phase is described within the
Dirac-Brueckner-Hartree-Fock (DBHF) approach \cite{DaFuFae05}.
Both pure NSs and those with a quark matter core
are consistent with modern CS constraints, see Fig.~1.
A lowering of the transition density results from increasing the 
diquark coupling strength, without affecting the stiffness, i.e.
the maximum mass.
Note that the transition to the color-favor-locking (CFL) phase renders the
hybrid star unstable \cite{Klahn:2006iw,Baldo:2002ju}.

Since the transition from the nuclear to these stiff QM EsoS 
is almost a crossover with only a tiny density jump, the resulting hybrid 
EoS is barely distinguishable from a purely hadronic one.
The corresponding hybrid stars ``masquerade'' as ordinary neutron stars 
 \cite{Alford:2004pf} regarding their masses, radii and similar observables of
compactness. ``Unmasking'' the NS interior could be possible by
analysing the cooling behavior \cite{Grigorian:2006pu} 
which eventually discriminates
nuclear from quark matter interiors due to the role of the pairing gaps
\cite{Blaschke:1999qx,Blaschke:2000dy,Popov:2005xa}.

\begin{center}
\begin{figure}[th]
\centerline{
\psfig{figure=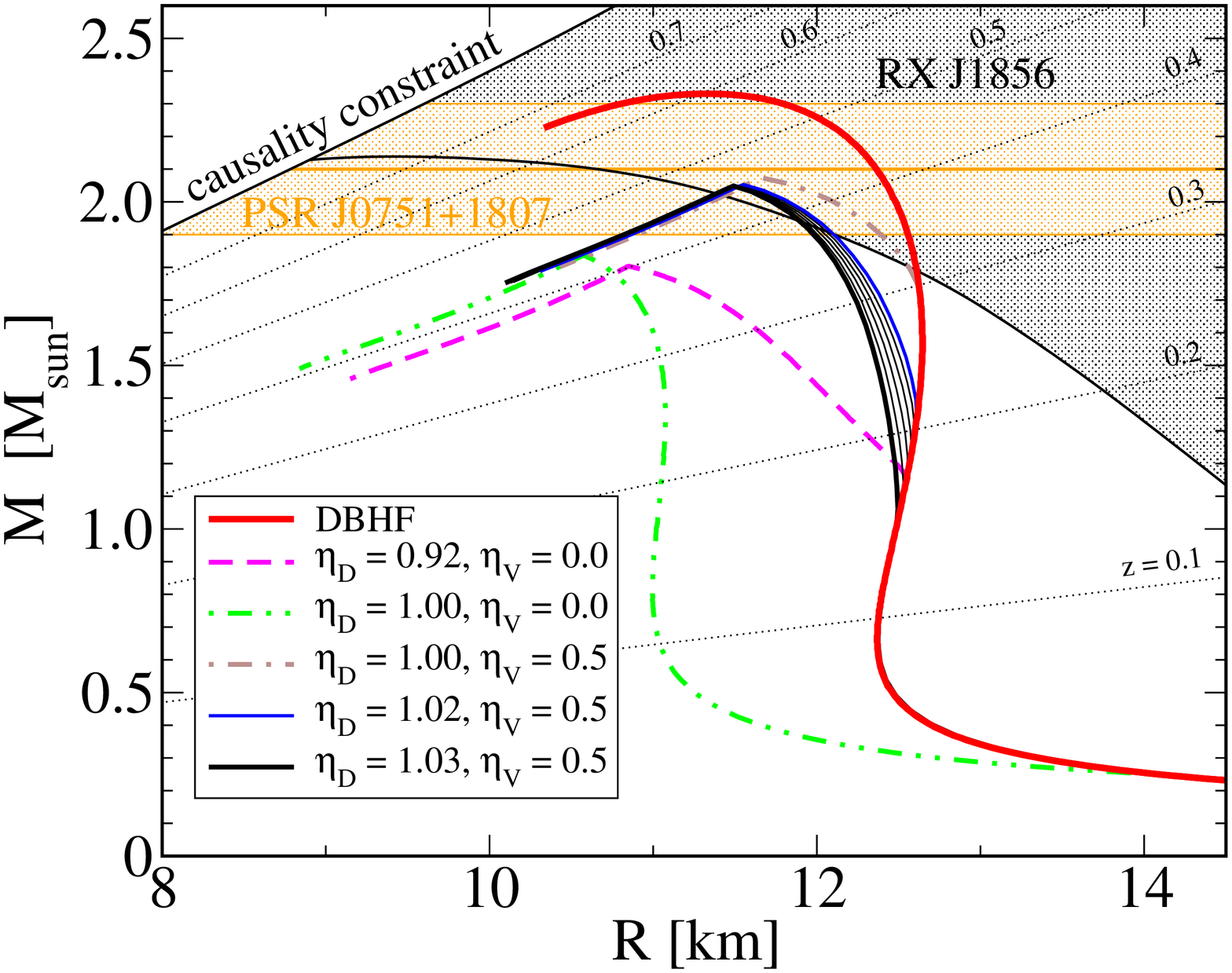,height=9.5cm,width=7.5cm,angle=-90}
\psfig{figure=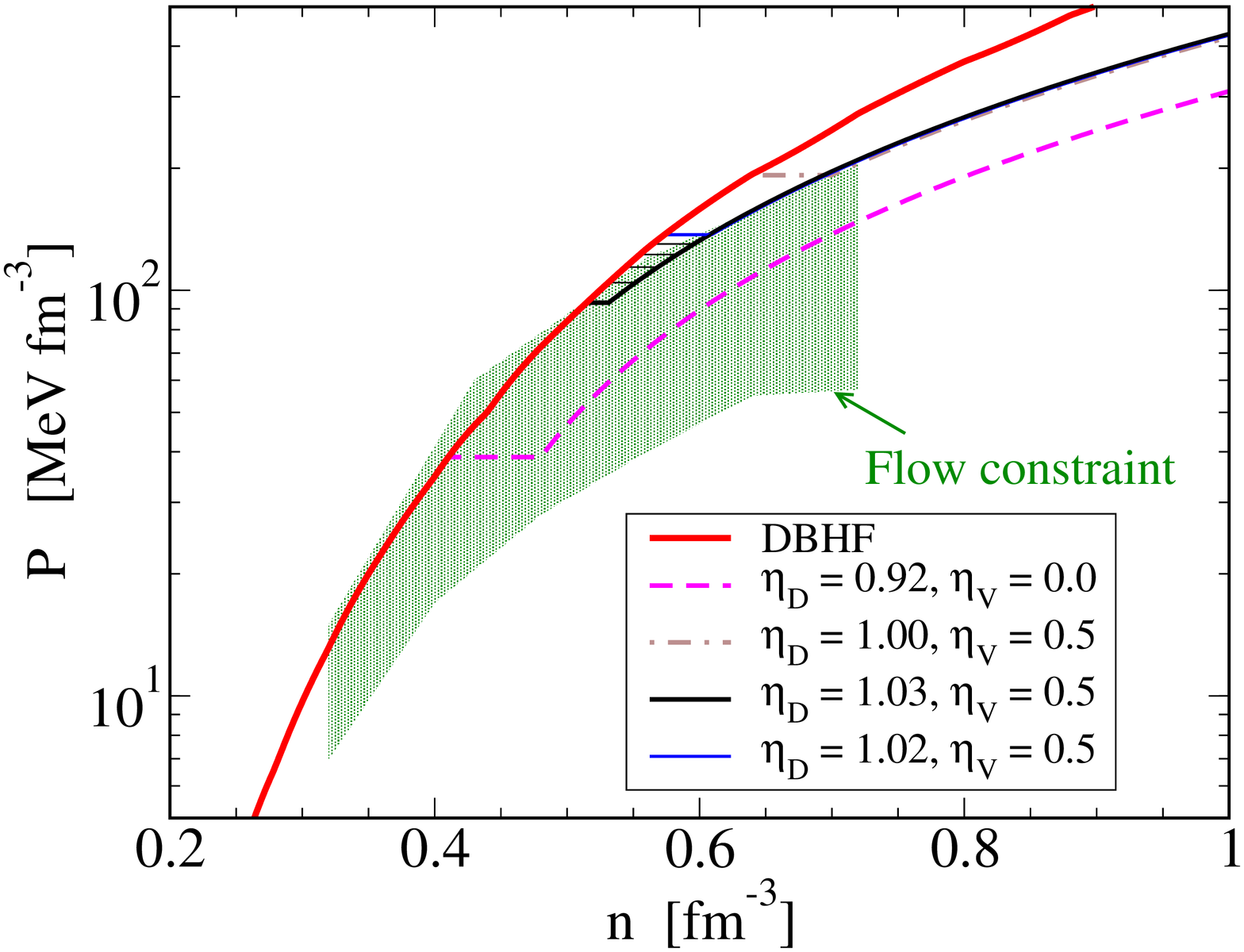,height=9.5cm,width=7.5cm,angle=-90}
}
\caption{{\small Hybrid EoS stiff enough to fulfil mass-radius constraints 
from PSR J0751+1807 and RX J18056 (left panel) but also soft enough to satisfy
Danielewicz' flow constraint (right).
A large vector meson coupling ($\eta_V$) increases the stiffness at high 
densities and thus the maximum mass, whereas increasing the diquark coupling 
($\eta_D$)lowers the onset of the deconfinement transition density and thus 
the critical star mass.} 
\label{fig:1}} 
\end{figure}
\end{center}


\vspace{-1cm}

\subsection*{Acknowledgements}
I acknowledge fruitful collaboration and discussions with many colleagues, 
in particular to the coauthors of \cite{Klahn:2006ir} as well as M. Alford,
A. Drago, S. Popov and F. Sandin.
This work has been partially supported by the Virtual Institute VH-VI-041 
of the Helmholtz Association 
and by GSI Darmstadt. I am grateful to Professor Faessler 
for his support of this project and to the Deutsche Forschungsgemeinschaft 
for a grant to participate at the Erice School.



\end{document}